# Dynamical Models of Stock Prices Based on Technical Trading Rules
# Part II: Analysis of the Model

Li-Xin Wang

*Abstract*—In Part II of the paper, we concentrate our analysis on the price dynamical model with the moving average rules developed in Part I of this paper. By decomposing the excessive demand function, we reveal that it is the interplay between trend-following and contrarian actions that generates the price chaos, and give parameter ranges for the price series to change from divergence to chaos and to oscillation. We prove that the price dynamical model has an infinite number of equilibriums, but all these equilibriums are unstable. We demonstrate the short-term predictability of the price volatility and derive the detailed formulas of the Lyapunov exponent as functions of the model parameters. We show that although the price is chaotic, the volatility converges to some constant very quickly at the rate of the Lyapunov exponent. We extract the formula relating the converged volatility to the model parameters based on Monte-Carlo simulations. We explore the circumstances under which the returns are uncorrelated and illustrate in details of how the correlation index changes with the model parameters. Finally, we plot the strange attractor and the return distribution of the chaotic price series to illustrate the complex structure and the fat-tailed distribution of the returns.

## I. INTRODUCTION

The analysis of the price dynamical models developed in Part I of this paper should aim at not only the properties of the models, but also the meanings of these properties in terms of financial economics. Specifically, we will show how the price dynamics of the models contribute to our understanding of four fundamental issues in financial economics: equilibrium, volatility, return predictability, and return independency.

Equilibrium is a fundamental idea in modern finance [6], [10], [19], [35], [41]. For example, the two core models in Modern Portfolio Theory [17] --- the Capital Asset Pricing Model (CAPM, the 1990 Nobel Prize winning model) and the Arbitrage Pricing Theory (APT) --- are based on the assumption that the prices will converge to the equilibrium point. However, the controversy around the concept of equilibrium has never ended in economics [40]. On one hand, the key results in general equilibrium theory --- the two theorems proved by Arrow and Debreu [3] --- are widely cited as providing the rigorous theoretical version of Adam Smith's invisible hand and demonstrating the desirable properties of a competitive economy; on the other hand, it is clear that the equilibrium in the general equilibrium theory is neither unique nor stable, meaning that there is no guarantee for the competitive market to converge to the desired equilibrium [1], [10], [27]. Based on the price dynamical models developed in this paper, the equilibrium-related questions can be addressed from a new angle. In particular, we will prove that the moving-average-rule-based price dynamical model has an infinite number of equilibriums, but all these equilibriums are unstable. These results are consistent with the main conclusions of the general equilibrium theory, albeit the general equilibrium theory is based on utility optimization whereas our price dynamical models come from the technical trading rules.

Another central concept in modern finance is volatility --- the standard deviation of the returns. The importance of volatility stems from two facts: (i) compared with the wilderness of returns, the volatility is much more stable so that reliable estimates of volatility based on real price data may be obtained, and (ii) volatility turns out to be the central variable in many core disciplines of modern finance such as asset pricing [24], portfolio allocation [17] and risk management [5], [9], [32]. There is a huge literature on volatility modeling [2], with the Nobel Prize winning ARCH model [18] and its generalized GARCH model [7] as stars. Many agent-based models [22], [43] were also proposed that can reproduce the empirical phenomena such as volatility clustering and excessive volatility. Most of these models are complex and the stochastic elements introduced in these models make it difficult to pinpoint the causes for these phenomena [29]. The contribution of this paper is to show that volatility is a fixed function of the model parameters that have clear physical meanings, therefore the causes for the phenomena can be precisely determined. For example, volatility clustering is caused by the clustered actions of traders which, in our price dynamical models, means that the strength parameters $a_i(t)$ are large in some time intervals and small in others. Similarly, excessive volatility is due to the strong actions (large strength parameters $a_i(t)$ in our models) of the traders at some important time points (such as when panic is spreading across the market and the chain firings of the stop-loss orders [26] result in very large strength parameters $a_i(t)$ in our models).

Foreseeing future returns is the dream of investors. A lot of research efforts have been undertaken to study whether financial indicators such as the earning-price ratios, dividend-price ratios, interest rates, corporate payout, etc., have predictive power for future returns (see [38] for a recent survey). The conclusions of these researches are confusing. On one hand, it was concluded in an influential paper [21] that "... these models have predicted poorly both in-sample and out-of-sample for 30 years now, ..., the profession has yet to find some variable that has meaningful and robust empirical equity premium forecasting power." On the other hand, the conclusion of the feature article [38] in the Handbook of Economic Forecasting [16] is that "the take-away message of this chapter is that methods are available for reliably improving stock return forecasts in an economically meaningful manner. Consequently, investors who account for stock return predictability with available forecasting procedures significantly outperform those who treat returns as entirely unpredictable." In this paper, we will study stock return predictability from a different angle --- through the price dynamical models. Since our technical-trading-rule-based price dynamical models are purely deterministic, short-term prediction is indeed possible with the "prediction horizon" characterized by the Lyapunov exponent which, as we will prove, is a fixed function of the model parameters.

Return independence is the key assumption in the random walk model which is the foundation of stochastic finance [9], [41]. Since real stock prices exhibit higher-order and nonlinear correlations [13], meaning that the price returns are in general not independent, the classical approach to deal with this problem is to model the volatility parameter in the random walk model as a random process (e.g. the ARCH and GARCH models). These classical models are complex (nonlinear stochastic equations) and since they are descriptive in nature (do not model directly the operations of traders), they could not provide quantitative links between return independence and trader actions. In this paper, we will show in detail how the returns generated by our price dynamical model are changing from positively correlated to uncorrelated and then to negatively correlated as the model parameters change. Since the model parameters have clear physical meanings such as the strength of the technical traders, our price dynamical models provide the detailed quantitative cause-effect links from trader actions to return correlation.

Part II of the paper is organized as follows. In Section II, we will analyze how chaos is generated within the price dynamical model and determine the parameter ranges for the price series to change from divergence to chaos and to oscillation. In Section III, we will prove mathematically that there are an infinite number of equilibriums for the price dynamical model, but all these equilibriums are unstable. In Section IV, we will illustrate the short-term predictability of the price volatility. In Section V, the Lyapunov exponent of the chaotic model will be determined and mathematical formulas of the Lyapunov exponent as functions as the model parameters will be derived. In Section VI, we will demonstrate the convergence of the volatility and extract a formula relating the converged volatility to the model parameters. In Section VII, we will study the correlations of the returns and illustrate how the correlation index changes with the model parameters. In Section VIII, we will plot the phase portrait and the distribution of the returns generated by the price dynamical model to illustrate the complex strange attractor and the fat-tailed return distribution. Finally, a few concluding remarks will be drawn in Section IX.

## II. How is the Chaos Generated: the Interplay between Trend-followers and Contrarians

Consider the price dynamic model driven by Heuristic 1 (Rule-1-Group) in Part I of this paper:

$$\ln(p_{t+1}) = \ln(p_t) + a_1 \, ed_1\bigl(x_t^{(m,n)}\bigr) \qquad (1)$$

where

$$x_t^{(m,n)} = \ln\left(\bar{p}_{t,m} \big/ \bar{p}_{t,n}\right), \quad \bar{p}_{t,n} = \frac{1}{n}\sum_{i=0}^{n-1} p_{t-i} \qquad (2)$$

is the log-ratio (relative change) of the price moving average of length-$m$ to the price moving average of length-$n$ with $m<n$, and

$$ed_1\bigl(x_t^{(m,n)}\bigr) = \frac{\sum_{i=1}^{7} c_i \mu_{A_i}\bigl(x_t^{(m,n)}\bigr)}{\sum_{i=1}^{7} \mu_{A_i}\bigl(x_t^{(m,n)}\bigr)} \qquad (3)$$

is the fuzzy system constructed from the seven fuzzy IF-THEN rules in Rule-1-Group, where $A_1 = PS$, $A_2 = PM$, $A_3 = PL$, $A_4 = NS$, $A_5 = NM$, $A_6 = NL$, $A_7 = AZ$ are the fuzzy sets whose membership functions are given in Fig. 1 of Part I of this paper, and $c_1 = 0.1$, $c_2 = 0.4$, $c_3 = -0.2$, $c_4 = -0.1$, $c_5 = -0.4$, $c_6 = 0.2$, $c_7 = 0$ are the centers of the fuzzy sets BS, BB,

SM, SS, SB, BM and AZ shown in Fig. 2 of Part I of this paper. Substituting these membership functions into (3), we obtain the detailed formula of $ed_1(x_t^{(m,n)})$ as follows:

$$ed_1(x_t^{(m,n)}) = \begin{cases} 0.2 & x_t^{(m,n)} \leq -3w \\ -\dfrac{0.6 x_t^{(m,n)}}{w} - 1.6 & -3w \leq x_t^{(m,n)} \leq -2w \\ \dfrac{0.3 x_t^{(m,n)}}{w} + 0.2 & -2w \leq x_t^{(m,n)} \leq -w \\ \dfrac{0.1 x_t^{(m,n)}}{w} & -w \leq x_t^{(m,n)} \leq w \\ \dfrac{0.3 x_t^{(m,n)}}{w} - 0.2 & w \leq x_t^{(m,n)} \leq 2w \\ -\dfrac{0.6 x_t^{(m,n)}}{w} + 1.6 & 2w \leq x_t^{(m,n)} \leq 3w \\ -0.2 & x_t^{(m,n)} \geq 3w \end{cases} \quad (4)$$

(See [44], [45] for the decomposition and approximation foundations of fuzzy systems) Fig. 1 plots $a_1 ed_1(x_t^{(m,n)})$. The task of this section is to analyze how and when chaos occurs with the excess demand $a_1 ed_1(x_t^{(m,n)})$ in Fig. 1.

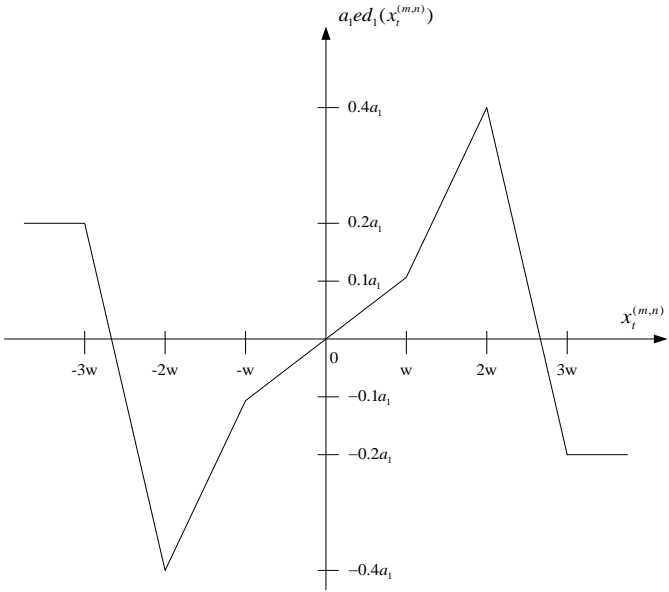

Fig. 1: Excess demand function $ed_1(x_t^{(m,n)})$ from Rule-1-Group traders.

From Fig. 1 we see that the excess demand function $a_1 ed_1(x_t^{(m,n)})$ intersects with the horizontal axis at the intervals [-3w,-2w] and [2w,3w], and from (4) we can compute that the intersected points are -2.66w and 2.66w, respectively. When $x_t^{(m,n)}$ is between $-2.66w$ and $2.66w$, the excess demand $a_1 ed_1(x_t^{(m,n)})$ has the same sign as $x_t^{(m,n)}$ so that the trend (raising or declining) continues; that is, in this case the trend-followers dominate the trading. When $x_t^{(m,n)}$ is smaller than $-2.66w$ or larger than $2.66w$, the excess demand $a_1 ed_1(x_t^{(m,n)})$ is in opposite sign of $x_t^{(m,n)}$ so that the trend begins to reverse; that is, in these cases contrarians have an upper hand. We show next that the interplay between trend-followers and contrarians generates the chaotic price series.

From (4) we see that there are four free parameters in the excess demand $a_1 ed_1(x_t^{(m,n)})$: $m$, $n$, $w$ and $a_1$, and they all have clear physical meanings: $m$, $n$ are the lengths of the shorter and longer moving averages, $w$ is the reference point when the traders say "Small", "Medium" or "Large" in their fuzzy trading rules ("small" means around $w$, "Medium" means around $2w$, and "Large" means around and larger than $3w$, as characterized by the membership functions shown in Fig. 1 of Part I of this paper), and $a_1$ is the relative strength of the traders using the trading rules with parameters $m$, $n$ and $w$. In other words, $m$, $n$ and $w$ are structural parameters that determine what kind of traders they are, and $a_1$ is the relative strength of this type of traders in action. The $w$ can also be interpreted as the "frequency" parameter because smaller (larger) $w$ implies more (less) frequent interchanges between trend-following and contrarian strategies. In our following analysis in this section, we fix the three structural parameters $m$, $n$ and $w$, and let $a_1$ free to change. Specifically, we choose $(m,n)=(1,5)$ and $w=0.01$ (1%).

Now we analyze what happens when $a_1$ takes values from small to large. Suppose the price $p_t$ is at a fixed value $p_{-1}$ before time zero ($p_t = p_{-1}$ for $t < 0$) and at $t = 0$ there is a price jump of $100 r_0$ percent ($p_0 = p_{-1}(1 + r_0)$). Suppose the initial price jump $r_0$ is not too large such that the initial $x_t^{(m,n)}$ is in the trend-following zone $(-2.66w, 2.66w)$ (see Fig. 1). If $a_1 \ (> 0)$ is very small, then the price change $\ln(p_{t+1}/p_t) = a_1 \, ed_1(x_t^{(m,n)})$ will be very small, so that the price trend will continue for a long time. Will the trend continue forever? Yes for small $a_1$ because as long as the $x_t^{(m,n)}$ remains in the trend-following zone $(-2.66w, 2.66w)$ during the process, the price $p_t$ will converge to some value (see Section III below for more discussion on this point). As $a_1$ increases, the price change $\ln(p_{t+1}/p_t) = a_1 \, ed_1(x_t^{(m,n)})$ is getting larger and larger to a point where the $x_t^{(m,n)}$ enters the contrarian zone $(2.66w, \infty)$ and the trend is reversed. If $a_1$ is not too large, the contrarians will draw the $x_t^{(m,n)}$ back to the trend-following zone and the trend-followers will once again push the $x_t^{(m,n)}$ to the contrarian zone; these back-and-forth actions generate chaos. When $a_1$ takes very large values, the price change $\ln(p_{t+1}/p_t) = a_1 \, ed_1(x_t^{(m,n)})$ is so large that the $x_t^{(m,n)}$ is pushed back and forth between the two contrarian zones $(-\infty, -2.66w)$ and $(2.66w, \infty)$; this causes oscillation. Fig. 2 shows a typical case of these three types of price trajectories: convergence ($a_1 = 0.049$), chaos ($a_1 = 0.26$), and oscillation ($a_1 = 0.39$), where $p_{-1} = 10$, $r_0 = w = 0.01$ and $(m, n) = (1,5)$.

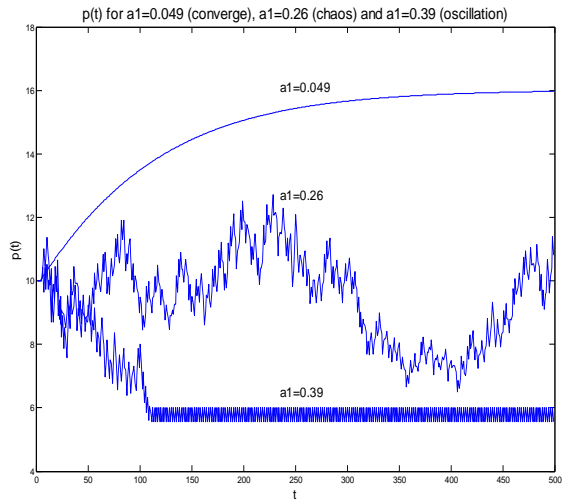

Fig. 2: Three price trajectories generated by (1) with $a_1 = 0.049$, 0.26 and 0.39, respectively, and $(m, n) = (1,5)$, $w = r_0 = 0.01$, $p_{-1} = 10$.

A natural question is: what are the ranges of the parameter $a_1$ for the price dynamics to be divergent, chaotic or oscillating? General theoretical results are not available at this point due to the complex high-dimensional nonlinearities in the price dynamical model (1)-(4). Through extensive simulations, we obtain the convergence, chaos and oscillation zones for $a_1$ shown in Fig. 3. We see in Fig. 3 there are two divergent zones when $a_1$ is changing from the convergent zone to the chaotic zone and from the chaotic zone to the oscillation zone, here divergence means the price keeps increasing to very large value or decreasing to zero so that the model does not represent any meaningful price series.

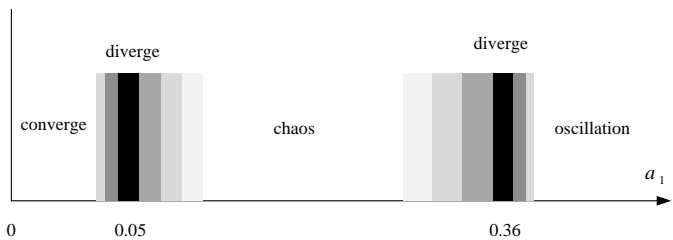

Fig. 3: Ranges of $a_1$ for the price trajectories to be convergent, divergent, chaotic or oscillating.

If we view convergence, chaos and oscillation as three stable states of the prices (making the analogy of the solid, liquid and gas states of matters), it is interesting to see what happen to the prices during the transition phases from convergence to chaos and from chaos to oscillation. Simulation results show that the price series exhibit some dramatic changes in the transition phases when the parameter $a_1$ changes only slightly. Specifically, Fig. 4 shows five price trajectories when $a_1$ is about to leave the convergent zone ($a_1 = 0.049, 0.0495, 0.0496, 0.0497$ and $0.0498$). We see from Fig. 4 that when $a_1$ changes from 0.0497 to 0.0498 --- a relative change of only $(0.0498 - 0.0497)/0.0497 \approx 0.002 = 0.2\%$, the convergent prices change from roughly 41 to 72 --- a relative change of $(72 - 41)/41 \approx 0.75 = 75\%$. Fig. 5 shows four price trajectories during the process when $a_1$ enters the oscillation zone ($a_1 = 0.365, 0.38, 0.39$ and $0.4$). We see from Fig. 5 that for $a_1 = 0.365$ the prices diverge to zero; with $a_1$ being increased to 0.38 the prices diverge to zero in a much slower and oscillation fashion; when $a_1$ is further increased to 0.39, the prices oscillate around some value (5.8); finally when $a_1$ gets large to 0.4, the prices oscillate around the initial condition $p_{-1} = 10$.

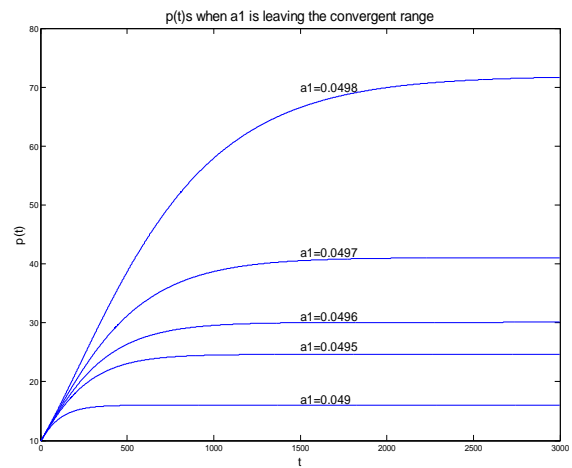

Fig. 4: The price trajectories for five different $a_1$'s when $a_1$ is leaving the convergent range.

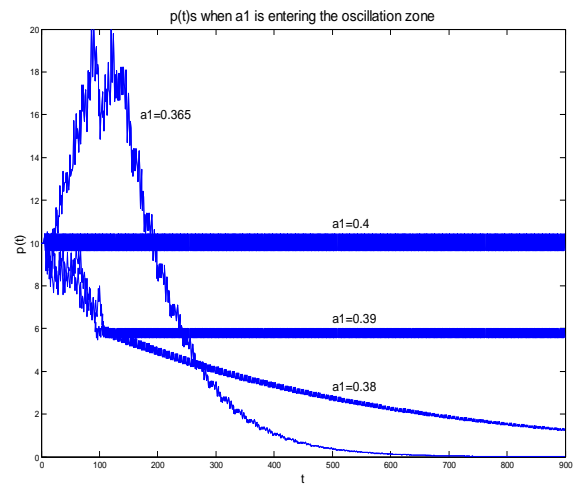

Fig. 5: The price trajectories for four different $a_1$'s when $a_1$ is entering the oscillation zone.

## III. ALL EQUILIBRIUMS ARE UNSTABLE

In this section we determine the equilibriums of the price dynamical model (1) and prove that all the equilibrium points are unstable. First, we recall the definition of equilibrium and stability (see for example [15]). Consider the dynamical system

$$y(t+1) = F[y(t)] \tag{5}$$

where $y(t) \in R^n$. A point $y^* \in R^n$ is *an equilibrium* of (5) if $y^* = F(y^*)$, so that if $y(t) = y^*$ for some $t$ then $y(t') = y^*$ for all $t' \geq t$. An equilibriums $y^*$ of (5) is *stable* if for any given $\varepsilon > 0$ and $t_1 > t_0$ there exists $\delta = \delta(\varepsilon, t_1)$ such that $||y(t_0) - y*|| < \delta$ implies $||yt - y*|| < \varepsilon$ for all $t \geq t1$. The equilibrium $y*$ is *asymptotically stable* if it is stable and $\lim_{t \to \infty} y(t) = y^*$. To find the equilibriums of the price dynamical model (1) and to study their stability, we first convert (1) into the form of (5).

Consider the price dynamical model (1). Defining $y_1(t) = p_{t-n+1}$, $y_2(t) = p_{t-n+2}$, ..., $y_n(t) = p_t$, we have from (1) that

$$\begin{cases} y_1(t+1) = y_2(t) \\ y_2(t+1) = y_3(t) \\ \quad \vdots \\ y_{n-1}(t+1) = y_n(t) \\ y_n(t+1) = y_n(t)e^{a_1\,ed_1\left(x_t^{(m,n)}\right)} \end{cases} \tag{6}$$

where $ed_1\left(x_t^{(m,n)}\right)$ is the fuzzy system (3) and (4) shown in Fig. 1,

$$x_t^{(m,n)} = \ln\left(\bar{p}_{t,m}/\bar{p}_{t,n}\right)$$

$$= \ln\left(\frac{1}{m}\sum_{i=n-m+1}^{n} y_i(t)\right) - \ln\left(\frac{1}{n}\sum_{i=1}^{n} y_i(t)\right) \tag{7}$$

and $m < n$. Let $y(t) = [y_1(t), y_2(t), ..., y_n(t)]^T$, then (6) becomes (5) with

$$F[y(t)] = \begin{bmatrix} y_2(t) \\ y_3(t) \\ \vdots \\ y_n(t) \\ y_n(t)e^{a_1\,ed_1\left(x_t^{(m,n)}\right)} \end{bmatrix} \tag{8}$$

That is, the price dynamical model (1) becomes (5) with $F[y(t)]$ given by (8). The following theorem gives the equilibriums of this model.

**Theorem 1:** For any positive number $p^* \in R_+$, $y^* = (p^*, ..., p^*)^T_{n \times 1}$ is an equilibrium of the price dynamical model (5) with $F[y(t)]$ given by (8).

Proof: Let $y(t) = y^* = (p^*, ..., p^*)^T_{n \times 1}$ for some $t$, we need to show $y(t+1) = F[y(t)] = F[y^*] = y^*$. Since $y(t) = [y_1(t), y_2(t), ..., y_n(t)]^T = (p^*, ..., p^*)^T_{n \times 1}$ implies $y_i(t) = p^*$ for $i=1,2,...,n$, in (7) we have $\frac{1}{m}\sum_{i=n-m+1}^{n} y_i(t) = p^*$ and $\frac{1}{n}\sum_{i=1}^{n} y_i(t) = p^*$ so that $x_t^{(m,n)} = 0$. From (4) we have $ed_1\left(x_t^{(m,n)}\right) = 0$ when $x_t^{(m,n)} = 0$, thus $y_n(t)e^{a_1\,ed_1\left(x_t^{(m,n)}\right)} = y_n(t)$. Consequently, from (5) and (8) we have $y(t+1) = F[y(t)] = [y_2(t), y_3(t), ..., y_n(t), y_n(t)]^T = y^*$. ∎

Theorem 1 shows that any price $p^*$ (any positive number) can be an equilibrium if $n$ consecutive prices equal $p^*$. Therefore, there are an infinite number of equilibriums for the price dynamical model (1). The following theorem shows that all these equilibriums are unstable.

**Theorem 2:** All equilibriums $y^* = (p^*, ..., p^*)^T_{n \times 1} \in R_+^n$ of the price dynamical model (5) with $F[y(t)]$ given by (8) are unstable.

Proof: The linearized equation of (5) at the equilibrium point $y^* = (p^*, ..., p^*)^T_{n \times 1}$ is

$$y(t+1) = Ay(t) \tag{9}$$

where $A$ is the Jacobian of $F[y(t)]$ at $y^*$, and from (8) we have

$$A = \begin{pmatrix} 0 & 1 & 0 & \cdots & 0 \\ 0 & 0 & 1 & \cdots & 0 \\ \vdots & \vdots & \vdots & \ddots & \vdots \\ 0 & 0 & 0 & \cdots & 1 \\ \frac{\partial f}{\partial y_1(t)}|_{y^*} & \frac{\partial f}{\partial y_2(t)}|_{y^*} & \frac{\partial f}{\partial y_3(t)}|_{y^*} & \cdots & \frac{\partial f}{\partial y_n(t)}|_{y^*} \end{pmatrix} \tag{10}$$

where $f$ is the $n$'th element of $F[y(t)]$:

$$f = y_n(t)e^{a_1\,ed_1\left(x_t^{(m,n)}\right)} \tag{11}$$

Let $\lambda_i$ ($i=1,2,...,n$) be the eigenvalues of A, then we have from (10) that

$$\prod_{i=1}^{n}(\lambda - \lambda_i) = |\lambda I - A|$$

$$= \lambda^n - \frac{\partial f}{\partial y_1(t)}|_{y^*}\lambda^{n-1} - \frac{\partial f}{\partial y_2(t)}|_{y^*}\lambda^{n-2} - \cdots - \frac{\partial f}{\partial y_n(t)}|_{y^*} \tag{12}$$

Hence $\left|\prod_{i=1}^{n} \lambda_i\right| = \left|\frac{\partial f}{\partial y_n(t)}\right|_{y^*}$. The standard Linearized Stability Theorem (see for example Theorem 5.15 of [15]) says that if at least one of the eigenvalues $\lambda_i$ ($i = 1,2,...,n$) is outside of the unit disk in the complex plane, then the equilibrium $y^*$ is unstable. Therefore if we can show $\left|\prod_{i=1}^{n} \lambda_i\right| = \left|\frac{\partial f}{\partial y_n(t)}\right|_{y^*} > 1$, then at least one of the $\lambda_i$'s must be outside of the unit disk and the equilibrium $y^*$ is unstable. From (11), (7) and (4), and noticing that $ed_1(x_t^{(m,n)}) = 0$ and $x_t^{(m,n)} = 0$ at the equilibrium $y^*$, we have, using the chain-rule, that

$$\frac{\partial f}{\partial y_n(t)}\bigg|_{y^*} =$$

$$\left(e^{a_1 ed_1(x_t^{(m,n)})} + a_1 y_n(t) e^{a_1 ed_1(x_t^{(m,n)})} \left(\frac{\partial ed_1(x_t^{(m,n)})}{\partial x_t^{(m,n)}}\right) \left(\frac{\partial x_t^{(m,n)}}{\partial y_n(t)}\right)\right)\bigg|_{y^*}$$

$$= 1 + a_1 p^* \left(\frac{0.1}{w}\right) \left(\frac{1}{m} - \frac{1}{n}\right) \quad (13)$$

Since $m<n$ and $a_1, p^*, w$ are positive, we have $\frac{\partial f}{\partial y_n(t)}\big|_{y^*} > 1$ and the theorem is proven. ■

As discussed in the Introduction, the general conclusion about the General Equilibrium Theory is that although the existence of equilibrium can be proved, neither uniqueness nor stability can be established for the equilibrium. Our results (Theorems 1 and 2) are consistent with these general conclusions: there exist an infinite number of equilibriums, and all these equilibriums are unstable. However, we saw in Section II that for small $a_1$ the prices converge to some fixed value (see for example Fig. 4). That is, instability of all the equilibrium points does not mean that the price generated by the model will not converge to some fixed value. This is an important observation because this unstable-but-convergent phenomenon seems to suggest that the classical concepts of stability --- developed for natural systems --- may not be suitable for social systems (such as stock markets). To make the arguments more clear, we consider a simpler linear return model as follows to illustrate the point.

Consider the simple trend-following or contrarian model:

$$\ln(p_{t+1}) = \ln(p_t) + a \left(\ln(p_t) - \ln(p_{t-1})\right) \quad (14)$$

where $a$ is a positive (trend-following) or negative (contrarian) number. Similar to Theorem 1, for any positive number $p^* \in R_+$, $y^* = (p^*, p^*)$ is an equilibrium point of (14) because once $p_{-1} = p_0 = p^*$, (14) guarantees $p_t = p^*$ for all $t > 0$. Now suppose that the prices stay at some equilibrium point $p^*$ before *t=0*, and there is a price jump of $100r_0$ percent at *t=0*: $\ln(p_0) = \ln(p^*) + r_0$. Then from (14) we can easily get

$$\lim_{t \to \infty} p_t = \lim_{t \to \infty} \left(p^* e^{\left(\frac{1-a^{t+1}}{1-a}\right) r_0}\right) = p^* e^{\left(\frac{1}{1-a}\right) r_0} \quad (15)$$

for $|a| < 1$. That is, for a disturbance $r_0$ at *t=0* the price will converge to a new equilibrium $p^* e^{\left(\frac{1}{1-a}\right) r_0}$ from the old equilibrium $p^*$ if $|a| < 1$. We see that the new converged value $p^* e^{\left(\frac{1}{1-a}\right) r_0}$ depends on all three variables: the old equilibrium $p^*$, the model parameter $a$ and the disturbance $r_0$. For $|a| < 1$ and $r_0 \neq 0$ we have $p^* e^{\left(\frac{1}{1-a}\right) r_0} \neq p^*$, which means the equilibrium $y^* = (p^*, p^*)$ cannot be asymptotically stable because any small disturbance $r_0$ will move the price away from the equilibrium forever. Hence we have similar conclusions for the simple trend-following model (14) as Theorems 1 and 2: the price model (14) has an infinite number of equilibriums $y^* = (p^*, p^*)$ for any $p^* \in R_+$ and all these equilibriums are unstable, although the prices always converge to a new equilibrium given in (15) if $|a| < 1$.

Since the classical stability concepts may not be suitable for the stock price models such as (1) and (14) as we discussed above, we may introduce a new stability concept, called *set-stability*, as follows: Let $y^*$ be an arbitrary equilibrium point of the dynamical system (5), if $y(t)$ converge to a new equilibrium $y^{**}$ after any small disturbance $\delta_0$ at time $t_0$ around $y^*$: $y(t_0) = y^* + \delta_0$, then the system (5) is said to be *set-stable*. According to this definition, the simple trend-following model (14) is set-stable if $|a| < 1$. For our price dynamical model (1) with $(m, n) = (1,5)$ and $w = 0.01$, extensive simulations (such as those shown in Fig. 4) suggest that it is set-stable if $|a_1| < 0.05$.

A standard approach in the agent-based price modeling literature [43] is to classify traders into two types: value investors who make investment decisions based on fundamentals, and trend followers who make investments in the direction of recent price movements. It is a common belief that value investors are rational and move the prices to their fundamental values, whereas trend followers are inherently destabilizing [33]. From our analysis above for the simple trend-following model (14) we see that this common opinion about trend followers is misleading. Indeed, trend followers push the price away from the old equilibrium, but the prices will converge to a new equilibrium as given in (15) as long as the strength of the trend-following actions is not too strong ($|a| < 1$). Although these old and new equilibriums are all unstable in the classical sense, the system as a whole is quite stable in ordinary times ($|a| < 1$) --- the prices simply move from one value to another in response to the evolving market conditions.

## IV. SHORT-TERM PREDICTABILITY: THE DIFFERENCE BETWEEN CHAOS AND RANDOM

The most colorful description of chaos is the Butterfly Effect --- a butterfly stirring the air in Hong Kong can transform storm systems in New York [20]. This is under the condition that Hong Kong and New York are very far away from each other. A butterfly stirring the air in Hong Kong cannot transform storms to its neighbors. In technical terms, the Butterfly Effect refers to the feature of chaotic systems that a small change in initial condition can result in very large changes as time moves forward. However, suppose we consider the situations only a few steps ahead from the initial time, the price behavior may be quite predictable.

Consider the price dynamical model (1)-(4) with initial condition $p_{-n+1} = \cdots = p_{-1} = y^*$ and $\ln(p_0) = \ln(y^*) + r_0$, where $y^*$ is an arbitrary positive number (the initial equilibrium price) and $r_0$ is the price disturbance at time zero. Let

$$r_t = \ln(p_t) - \ln(p_{t-1}) \approx \frac{p_t - p_{t-1}}{p_{t-1}} \quad (16)$$

be the returns generated by the price model (1). To visualize the dynamical evolution of the returns $r_t$, we choose the price disturbance $r_0$ to be a random variable and perform Monte-Carlo simulations. Specifically, let $r_0 = v_0 \varepsilon_0$ where $v_0$ is a positive constant and $\varepsilon_0$ is a Gaussian random variable with mean 0 and variance 1, and we ran the price dynamical model (1) with different realizations of $r_0$. With the parameters $(m,n)=(1,5)$, $w=0.01$, $a_1 = 0.17$ and $p^* = 10$, Fig. 6 shows the simulation results, where the top sub-figure plots the return trajectories $r_t$ of 100 runs with $v_0 = 10^{-5}$ (very small), the next two sub-figures show the same for $v_0 = 10^{-4}$ and $10^{-3}$, respectively, and (for comparison) the bottom sub-figure plots the returns $r_t$ of 100 simulation runs of the random walk model:

$$\ln(p_{t+1}) = \ln(p_t) + \sigma \varepsilon_t \quad (17)$$

with $p_{-1} = p^*$, $v_0 = 10^{-5}$ and $\sigma = 0.03$ for $t \geq 0$. We see from Fig. 6 the fundamental difference between chaos and random: the chaotic returns (top three sub-figures) change gradually from the initial values to the steady state, whereas the random walk returns (bottom sub-figure) reach the steady state in the first step without any transition period.

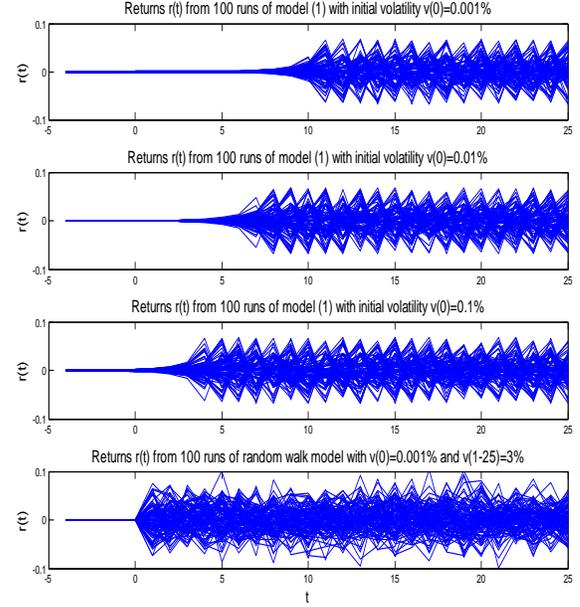

Fig. 6: Monte Carlo simulations (100 return trajectories in each sub-figure) of the price dynamical model (1) (top three sub-figures) and random walk model (17) (bottom sub-figure) with different initial conditions.

To make the picture clearer, we define *volatility at time t* as:

$$v(t) = \left( \frac{\sum_{j=1}^{S} \left( \ln(p_t^j) - \ln(p_{t-1}^j) \right)^2}{S} \right)^{1/2} \quad (18)$$

where $p_t^j$ is the price of the *j'th* simulation run and $S$ is the total number of the Monte Carlo simulations; i.e., $v(t)$ is the sample estimate of the standard deviation of the return $r_t = \ln(p_t/p_{t-1})$ at time *t* (the mean of $r_t$ is assumed to be zero for prices in the chaotic domain due to the symmetry of the excess demand function $ed_1(x_t^{(m,n)})$). For the same Monte Carlo simulations in Fig. 6, the corresponding $v(t)$'s are plotted in Fig. 7. We see from Fig. 7 that the volatility $v(t)$ of the chaotic model (1) increases gradually from the initial $v_0 = 0.001\%, 0.01\%$ or $0.1\%$ to some steady value (around 0.03), whereas the $v(t)$ of the random walk model (17) reaches the steady value $\sigma = 0.03$ immediately at the first time point *t*=1. Here again we see the difference between chaos and random: Price volatility from the chaotic model is short-term predictable, whereas price volatility of the random walk model is unpredictable.

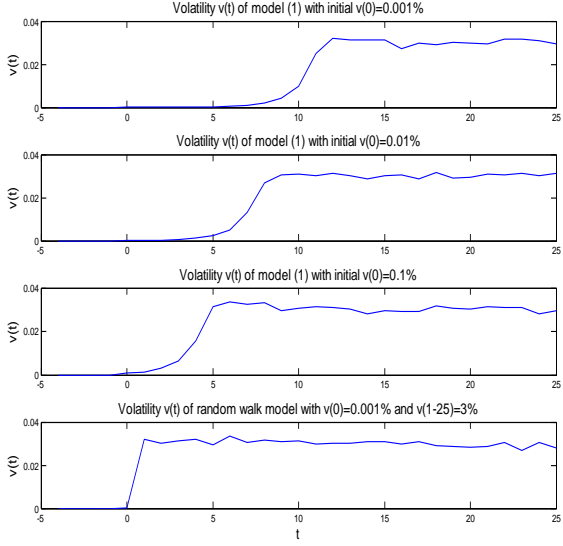

Fig. 7: The volatilities $v(t)$ computed from the Monte Carlo return trajectories in Fig. 6.

## V. LYAPUNOV EXPONENT

As discussed in the last section, the key feature of a chaotic system is sensitive dependence on initial conditions. In the chaos theory literature [28], [39], Lyapunov exponent is used to quantify this sensitive dependence. Consider two nearby initial conditions separated by a small quantity $\delta_0$, and let $\delta_t$ be the separation of the two trajectories after *t* steps. If $\delta_t \approx \delta_0 e^{Lt}$ for some positive constant *L*, then the *L* is called the *Lyapunov exponent* [39], [42]. Taking log on both sides of $\delta_t \approx \delta_0 e^{Lt}$ we have $\ln(\delta_t) \approx \ln(\delta_0) + Lt$, so that if we plot $\delta_t$ as a function of *t* in the log-t scale, the slope of the line is the Lyapunov exponent.

Let the variable of interest be the return $r_t = \ln(p_t) - \ln(p_{t-1})$ generated by the chaotic dynamic model (1)-(4) with initial condition $p_{-n+1} = \cdots = p_{-1} = y^*$ and $\ln(p_0) = \ln(y^*) + r_0$. Consider two nearby initial $r_0$ and $r_0'$, generating two return sequences $r_t$ and $r_t'$, respectively; if we can show that the returns $r_t$ in the first few time steps satisfies $r_t \approx r_0 e^{Lt}$ for some positive constant *L*, then the *L* is the Lyapunov exponent because $r_t \approx r_0 e^{Lt}$ together with $r_t' \approx r_0' e^{Lt}$ gives $|r_t - r_t'| \approx |r_0 - r_0'| e^{Lt}$ which implies that the separation of the two return trajectories $\delta_t \equiv |r_t - r_t'|$ satisfies $\delta_t \approx \delta_0 e^{Lt}$. Before we derive the mathematical formulas of the Lyapunov exponent *L* as functions of the model parameters, we perform some simulations to get a feeling of the Lyapunov exponent. According to definition (18), the volatility $v(t)$ is the estimate of the standard deviation of the return $r_t$ based on Monte-Carlo simulations, thus we use the volatility $v(t)$ as a representative for the return $r_t$ in the computing of the Lyapunov exponent; that is, if it can be shown that $v(t) \approx v_0 e^{Lt}$, then the *L* is the Lyapunov exponent. Therefore, if we plot $v(t)$ versus *t* in the log-t scale, then the slope of the line gives the Lyapunov exponent. Fig. 8 plots the same simulation results as Fig. 7 in the log-t scale (adding one more case with initial $v_0=10^{-6}$). By measuring the slopes of the lines in Fig. 8, we obtain the Lyapunov exponent roughly equal to 0.74 for this case (model (1) with parameters *(m,n)=(1,5)*, *w*=0.01 and $a_1 = 0.17$).

To see how the volatility changes for different values of $a_1$, we plot in Fig. 9 the volatilities *v(t)* in the log-t scale with initial $\sigma = 10^{-6}$ and $a_1$ taking 0.12, 0.17, 0.22, 0.27 and 0.32, respectively. From Fig. 9 we see that as $a_1$ increases, the Lyapunov exponents (slopes of the $\log(v(t))$ plots) are getting larger, meaning that the volatilities are settling down to the steady values faster. Since the physical meaning of the parameter $a_1$ is the trading strength of the Rule-1-Group traders, larger $a_1$ implies higher trading activity which results in faster convergence to the steady volatility.

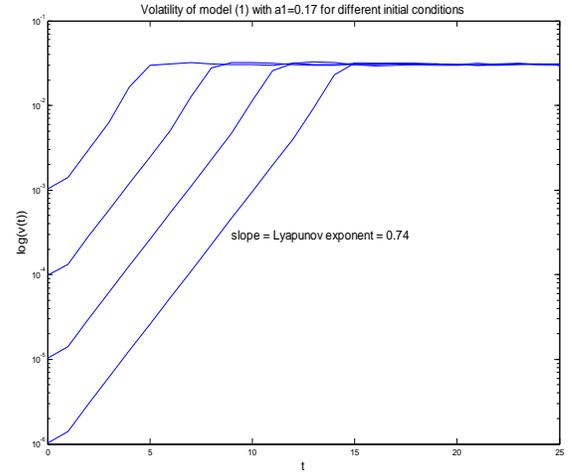

Fig. 8: The volatilities $v(t)$ computed from Monte Carlo simulations of the price dynamical model (1) with *(m,n)=(1,5)*, *w*=0.01 and $a_1 = 0.17$ for four different initial conditions: $v_0=10^{-3}$, $v_0=10^{-4}$, $v_0=10^{-5}$ and $v_0=10^{-6}$.

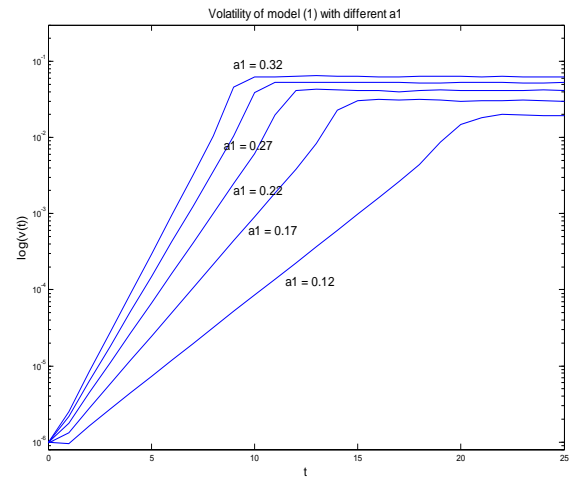

Fig. 9: The volatilities $v(t)$ computed from Monte Carlo simulations of the price dynamical model (1) with *(m,n)=(1,5)*, *w*=0.01, initial condition $v_0=10^{-6}$, and $a_1$ taking the five different values.

We now derive the mathematical formulas of the Lyapunov exponent $L$ as functions of the model parameters in the following lemma.

**Lemma 1:** Consider the price dynamical model (1)-(4) with the structural parameters *(m,n)=(1,5)*, *w=0.01* fixed and the strength parameter $a_1$ taking values in the chaotic range of Fig. 3 to generate chaotic price series. Let $L$ be the Lyapunov exponent of such system, then we have approximately that

$$L \approx \ln\left(\frac{3}{4} + 8a_1\right) \quad (19)$$

For the more general case of *m=1* and the other three parameters *n, w* and $a_1$ are free to change, the Lyapunov exponent is given approximately by

$$L \approx \ln\left(\frac{n-2}{n-1} + \frac{0.1 a_1}{w}\left(1 - \frac{1}{n}\right)\right) \quad (20)$$

Proof of this lemma is given in the Appendix.

## VI. VOLATILITY AS FUNCTION OF MODEL PARAMETERS

An important observation from the simulation results in Figs. 7 to 9 is that no matter what the initial conditions are, the volatilities $v(t)$ always converge to the same constant after a small number of steps, and this constant depends only on the model parameters. For the parameter setting of Fig. 8 (*(m,n)=(1,5)*, *w=0.01*, $a_1 = 0.17$), this constant is around 0.03. Although a general mathematical proof for the convergence of the volatility $v(t)$ to a constant as $t$ goes to infinity is not available at this point[1], the $v(t)$'s in all our simulations converged to some constants which depend only on the model parameters *m, n, w* and $a_1$, and are independent of the initial conditions $v_0$.

Fig. 10 shows the converged volatility $v_\infty = \lim_{t\to\infty} v(t)$ of the prices generated by model (1) as a function of the strength parameter $a_1$ for some fixed *w* and *(m,n)=(1,5)*. From Fig. 10 we see that when $a_1$ is small, the converged volatility is zero; this agrees with our analysis in Sections II and III that in this case only trend followers trade with weak activities so that the prices converge to some constant (see Fig. 4; the volatility of a converged price series is zero). As $a_1$ is getting larger to enter the chaotic zone, the converged volatility suddenly increases very rapidly. In the chaotic zone, the converged volatility $v_\infty$ shows complex behavior: first as a fast increasing function of $a_1$, then a slowly increasing function, and finally increases fast again as $a_1$ is entering the oscillation zone. In the oscillation zone, the converged volatility $v_\infty$ is a linearly increasing function of $a_1$.

[1] We need to prove, e.g., that the Frobenius-Perron operator [12], [14], [28] of model (1) has a unique fixed point that is reachable from any initial density.

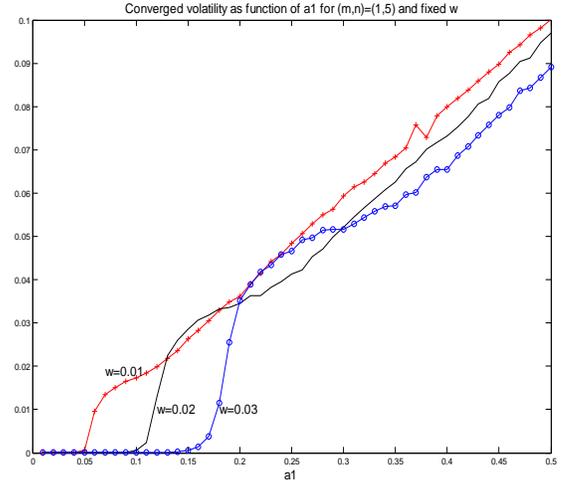

Fig. 10: Converged volatility $v_\infty$ as function of $a_1$ for some fixed *w* and *(m,n)=(1,5)*.

Similarly, Fig. 11 plots the converged volatility $v_\infty$ as a function of the frequency parameter *w* for some fixed $a_1$. We see from Fig. 11 that when *w* is very small, the price is in the oscillation zone and the $v_\infty$ does not change with *w*; this can be understood from the excess demand function $a_1 ed_1(x_t^{(m,n)})$ in Fig. 1 that when *w* is very small comparing to $a_1$, the $x_t^{(m,n)}$ is either larger than *3w* or smaller than *-3w* such that the returns $r_{t+1} = a_1 ed_1(x_t^{(m,n)})$ oscillate between $0.2a_1$ and $-0.2a_1$, which gives $v_\infty = 0.2a_1$. As *w* increases, the prices are entering the chaotic zone where smaller returns (comparing to the large returns $\pm 0.2a_1$ in the oscillation zone) are occurring more and more frequently, which results in smaller $v_\infty$. In the chaotic zone, the converged volatility $v_\infty$ decreases first as *w* is getting larger and increases again as *w* is approaching the convergent zone. During the transition from the chaotic zone to the convergent zone, the prices change violently and the result is a sharp decline of $v_\infty$ to zero.

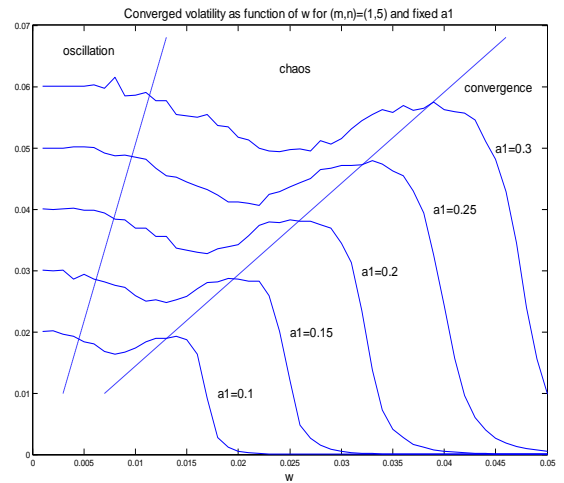

Fig. 11: Converged volatility $v_\infty$ as function of *w* for some fixed $a_1$ and *(m,n)=(1,5)*.

Based on extensive Monte Carlo simulations such as those in Figs. 10 and 11, we have the following result:

**Result 1:** Consider the price dynamical model (1)-(4) and let $v_\infty$ be the converged value of the volatility $v(t)$ defined in (18). For $(m,n)=(1,5)$ and $a_1, w$ in the chaos zone of Fig. 11, we have approximately that

$$v_\infty \approx 0.19 a_1 + 0.03 a_1 \sin\left[\frac{\pi}{0.1 a_1}(w + 0.06 a_1)\right] \quad (21)$$

which is obtained by fitting the curves in the chaos zone of Fig. 11 with a basic sin function. ∎

Since the volatility $v(t)$ usually converges to the steady value $v_\infty$ very quickly (see Figs. 7 to 9), we can in general ignore the transition period and view the $v_\infty$ computed from (21) as the volatility of the prices generated by the price dynamical model (1). An important stylized fact of real stock prices is volatility clustering (or volatility persistency), i.e., large (small) price changes are followed by other large (small) price changes [11], [13], [34]. Volatility clustering can be easily interpreted according to (21) as follows: since volatility is a fixed function of the trading strength parameter $a_1$ and the frequency parameter $w$, volatility persistency is simply the reflection of the slow time-varying nature of the model parameters $a_1$ and $w$ (as compared with the fast time-varying stock prices). Consider the scenario that a good news was announced for a company and people jumped in to buy the stock of this company. Clearly, the buy action would in general continue for a while when more people learned the news and prepared the money to buy the stock; this would keep the strength parameter $a_1$ around some large value for some time, and the result was volatility persistency.

Finally we prove a formula for the converged volatility when the model is in the oscillation mode.

**Lemma 2:** Consider the price dynamical model (1)-(4) with $m=1$ and the other three parameters $n$, $w$ and $a_1$ are free to change. Suppose that in the steady state the price $p_t$ oscillates between two fixed values, then the volatility of the steady state prices is given by

$$v_\infty = 0.2 a_1 \quad (22)$$

and the parameters $n, w, a_1$ satisfy the constraint:

$$\frac{1}{n}\text{int}\left(\frac{n}{2}\right) a_1 \geq 15w \quad (23)$$

where int is the take-the-integer operator.

Proof: With return $r_t$ defined in (16), model (1) yields $p_t = p^* e^{\sum_{i=0}^{t} r_i}$ where $p^*$ is the initial price. Using the approximate formula $1 + s \approx e^s$ for small $s$, we have for large $t$ that

$$x_t^{(1,n)} = \ln\left(\frac{p^* e^{\sum_{i=0}^{t} r_i}}{\frac{1}{n}\left(\sum_{j=1}^{n} p^* e^{\sum_{i=0}^{t-j+1} r_i}\right)}\right)$$

$$\approx \left(\frac{n-1}{n}\right) r_t + \left(\frac{n-2}{n}\right) r_{t-1} + \cdots + \left(\frac{1}{n}\right) r_{t-n+2} \quad (24)$$

($\sum_{i=0}^{t} r_i$ is small since returns are zero-mean.) Since the steady state prices oscillate between two fixed values, the returns $r_t, r_{t-1}, r_{t-2}, \ldots$ in the steady state must equal to $\pm v_\infty$ with alternative positive and negative signs, where $v_\infty$ is the steady state volatility. Let $r_t = v_\infty, r_{t-1} = -v_\infty, r_{t-2} = v_\infty, r_{t-3} = -v_\infty, \ldots$, then (24) gives $x_t^{(1,n)} \approx \frac{1}{n}\text{int}\left(\frac{n}{2}\right) v_\infty$. From Fig. 1 and (4) we see that in order for $r_{t+1} = -v_\infty$, we must have $x_t^{(1,n)} \approx \frac{1}{n}\text{int}\left(\frac{n}{2}\right) v_\infty \geq 3w$ that gives $r_{t+1} = a_1 ed_1(x_t^{(1,n)}) = -0.2 a_1 = -v_\infty$; this proves (22). Substituting $v_\infty = 0.2 a_1$ into the condition $\frac{1}{n}\text{int}\left(\frac{n}{2}\right) v_\infty \geq 3w$ yields (23). ∎

Lemma 2 shows that when $a_1$ is very large or $w$ is very small (such that (23) is satisfied), the steady state volatility $v_\infty$ depends only on the strength parameter $a_1$ (as given by (22)). This phenomenon is confirmed by the simulation results in Figs. 10 and 11: in Fig. 10 we see that as $a_1$ is getting larger, the three curves with different $w$'s are converging to the same line $v_\infty = 0.2\, a_1$; and Fig. 11 shows that the five curves for different $a_1$'s are horizontal lines (independent of $w$) when $w$ is very small, and the numbers in the figure agree with the formula $v_\infty = 0.2\, a_1$.

## VII. ARE THE RETURNS "UNCORRELATED"?

A fundamental assumption of the random walk model (17) is that the returns $\sigma \varepsilon_t$ must be independent. Now we ask: Are the returns $r_t = \ln(p_t/p_{t-1}) = a_1 ed_1(x_{t-1}^{(m,n)})$ generated by our chaotic model (1) "uncorrelated"? Because the returns $a_1 ed_1(x_{t-1}^{(m,n)})$ are deterministic, what does this "uncorrelated" means?

We know that if the returns $\sigma \varepsilon_t$ in the random walk model (17) are uncorrelated, then the standard deviation of $\ln(p_t/p_0)$ equals $\sigma\sqrt{t}$, i.e.,

$$\left(E\left\{\left(\ln(p_t/p_0)\right)^2\right\}\right)^{1/2} = \sigma\sqrt{t} \quad (25)$$

Therefore, if some price series $p_t$ satisfies (25) approximately, we can think of the returns from this price series as being uncorrelated. If we want to use (25) to check the correlation of the returns generated by model (1), the first question is how to compute the expectation $E\{*\}$ in (25). Since our price dynamical model (1) is deterministic, the returns generated by the model are not random, so what does the expectation of a non-random variable mean? We address this problem by making use of the Butterfly Effect of the chaotic systems. That is, we change the initial prices slightly and run the chaotic model many times, the price series so generated are viewed as different realizations of a random process. More specifically, we perform Monte Carlo simulations for the price dynamical model (1) with initial condition $p_{-n+1} = \cdots = p_{-1} = y^*$, $\ln(p_0) = \ln(y^*) + v_0 \varepsilon_0$ where $y^*, v_0$ are constants and $\varepsilon_0$ is a zero-mean unit-variance Gaussian random variable (same as we did for the simulations in Figs. 6 and 7), and then use the average over the different simulation runs for the expectation $E\{*\}$ in (25). Let $p_t^j$ be the price of the *j'th* simulation run and $S$ is the total number of the Monte Carlo simulations, define the *drift of log price in time t* as

$$d(t) = \left( \frac{\sum_{j=1}^{S} \left( \ln(p_t^j) - \ln(p_0^j) \right)^2}{S} \right)^{1/2} \quad (26)$$

Then we say the returns are *uncorrelated* if $d(t)$ is equal to $v_\infty \sqrt{t}$. Furthermore, define the *distance-to-uncorrelated* as

$$DU = \frac{\sum_{t=N_1}^{N_2} (d(t) - v_\infty \sqrt{t})}{N_2 - N_1 + 1} \quad (27)$$

where $v_\infty$ is the converged volatility and $N_2, N_1$ are some large numbers with $N_2 > N_1$.

The three sub-figures in Fig. 12 plot the drifts $d(t)$ computed from $S = 600$ simulation runs of model (1) and the corresponding random walk drifts $v_\infty \sqrt{t}$ for the cases of $a_1 = 0.12$ (top, with $v_\infty = 0.019$ computed from (21)), $a_1 = 0.14$ (middle, with $v_\infty = 0.023$ computed from (21)) and $a_1 = 0.18$ (bottom, with $v_\infty = 0.031$ computed from (21)), respectively (for all the cases $w = 0.01$ and *(m,n)=(1,5)*). We see from Fig. 12 that when $a_1$ is small (the top sub-figure of Fig. 12), the drift $d(t)$ is increasing faster than $v_\infty \sqrt{t}$ (super-diffusion in the language of econophysics [9], [11]) ; as $a_1$ increases (the middle sub-figure of Fig. 12), the drift $d(t)$ becomes very close to $v_\infty \sqrt{t}$ (norm-diffusion); finally when $a_1$ gets large (the bottom sub-figure of Fig. 12), the drift $d(t)$ is increasing slower than $v_\infty \sqrt{t}$ (sub-diffusion). The reason for the phenomena (super-diffusion, diffusion and sub-diffusion) in Fig. 12 is the following: When the strength $a_1$ is small relative to the $w$, the trend-followers have an upper hand so that the prices tend to move in the same direction (super-diffusion), which results in large drift; then, as $a_1$ increases, the prices $p_t$ become more chaotic and to a point where the chaos reaches the maximum (pure diffusion) such that the returns become uncorrelated ($d(t)$ is very close to $v_\infty \sqrt{t}$); finally, as $a_1$ increases furthermore, the contrarians are gaining an upper hand so that the prices $p_t$ tend to oscillate (sub-diffusion), which makes the drift small.

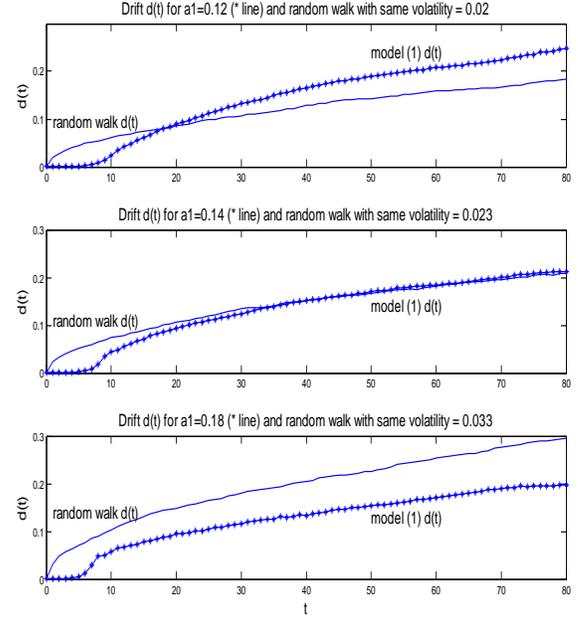

Fig. 12: The drifts $d(t)$ (26) of the price dynamical model (1) and the random walk model (17) with the same volatility for the cases of $a_1 = 0.12$ (top; super-diffusion), $a_1 = 0.14$ (middle; diffusion) and $a_1 = 0.18$ (bottom; sub-diffusion).

To see more details of how the correlations of the returns change with the model parameters, we plot in Fig. 13 the distance-to-uncorrelated DU defined in (27) as function of $a_1$ for some fixed *w* with *(m,n)=(1,5)*, $N_2 = 45, N_1 = 30$ and $S = 600$. Similarly, Fig. 14 plots the distance-to-uncorrelated as function of *w* for some fixed $a_1$ and *(m,n)=(1,5)*. We see from Fig. 13 that as $a_1$ increases from very small value, the DU first increases when the model is moving from the convergent zone to the chaotic zone. Then, as $a_1$ moves further into the chaotic zone, the prices become more and more chaotic such that the DU begins to decrease. The chaos reaches the maximum when the DU curves intersect with the zero line, and at these intersection points the DU equals zero and the returns are uncorrelated. As $a_1$ increases furthermore, the model is approaching the oscillation zone and the drift $d(t)$ increases slower than $v_\infty \sqrt{t}$, which results in negative DU. When $a_1$ is inside of the oscillation zone, the prices oscillate between some fixed values and the drift $d(t)$ stops increasing; in this case the converged volatility $v_\infty$ increases linearly with $a_1$ according to (22) of Lemma 2 so that the DU moves further into the negative territory, as demonstrated in Fig. 13. Fig. 14 can be interpreted in a similar fashion.

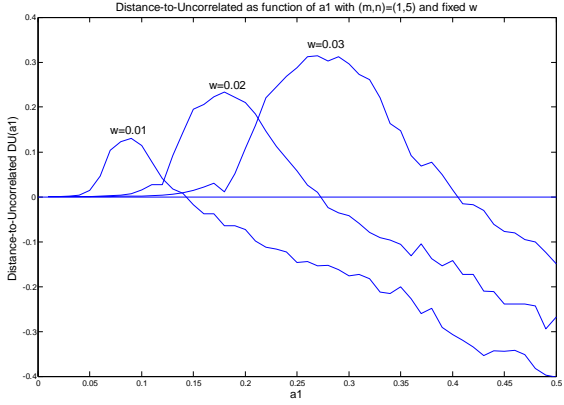

Fig. 13: The distance-to-uncorrelated (27) as function of $a_1$ for some fixed $w$ and $(m,n)=(1,5)$.

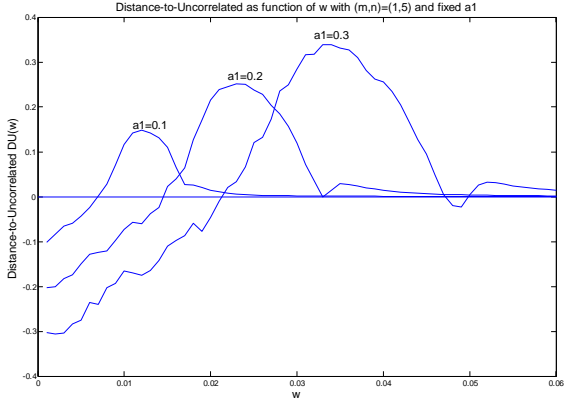

Fig. 14: The distance-to-uncorrelated (27) as function of $w$ for some fixed $a_1$ and $(m,n)=(1,5)$.

Based on extensive Monte Carlo simulations such as those in Figs. 13 and 14, we have the following result.

**Result 2:** Consider the returns $r_t = \ln(p_t/p_{t-1})$ generated by the price dynamical model (1) with $(m,n)=(1,5)$. If the strength parameter $a_1$ and the frequency parameter $w$ satisfy approximately the following linear relation:

$$a_1 \approx 14.28\, w \tag{28}$$

then the returns $r_t$ ($t = 1,2,3,...$) are uncorrelated in the sense that the distance-to-uncorrelated DU defined in (27) is approximately zero. ∎

In the study of return correlation, another important criterion is the auto-correlation of the returns. Fig. 15 shows the auto-correlations[2] $E_t\{r_t r_{t+i}\}$ of the returns generated by the price dynamical model (1) with $(m,n)=(1,5)$, $w=0.01$ and three different $a_1$'s: $a_1 = 0.1, 0.14$ and $0.34$ for the top, middle and bottom sub-figures, respectively. From Fig. 15 we see that when $a_1$ is small (the $a_1 = 0.1$ case), there is a persistent small positive correlation between the returns; this is due to the dominance of the trend-following actions when $a_1$ is small relative to $w$. As $a_1$ increases (the $a_1 = 0.14, 0.34$ cases), the auto-correlations decay to zero very quickly, confirming the chaotic nature of the prices generated by the model and also agreeing with the real stock prices [11], [13].

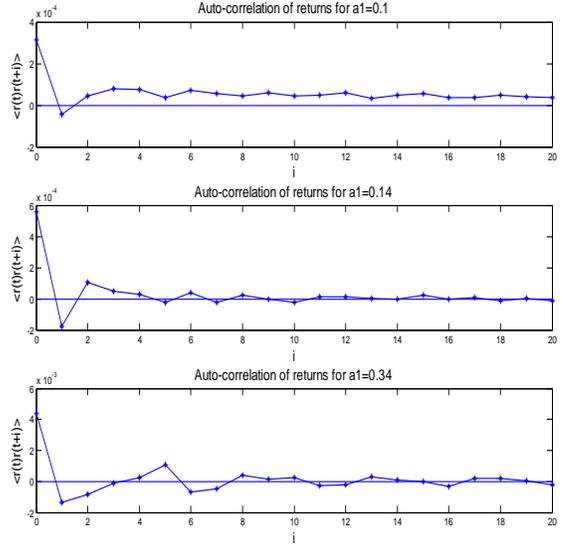

Fig. 15: The auto-correlations $E_t\{r_t r_{t+i}\}$ of the returns generated by the price dynamical model (1) with $(m,n)=(1,5)$, $w=0.01$ and $a_1 = 0.1$ (top), $a_1 = 0.14$ (middle), $a_1 = 0.34$ (bottom).

## VIII. STRANGE ATTRACTOR AND FAT-TAILED RETURN DISTRIBUTION

Phase portraits of chaotic systems, called strange attractors, are a useful way to illustrate the complexity and interesting structures of chaotic dynamics. Perhaps, the most lasting memory of the famous chaotic systems for an ordinary person may be their colorful strange attractors, such as the butterfly of the Lorenz attractor [31]. We now plot the phase portrait of our price dynamical model (1).

Fig. 16 plots the trajectory of a simulation run, in the 2D return subspace $r_{t-1}$-vs-$r_t$, of the price dynamical model (1) with $(m,n)=(1,5)$, $w = 0.01$ and $a_1 = 0.14$. Since the order of the system is $n=5$, the strange attractor in Fig. 16 is the projection of the phase portrait on the 2D subspace.

An important stylized fact of real stock prices is their fat-tailed distribution [11], i.e., the frequency of occurrences of large returns (positive or negative) is much higher than what predicted by the Gaussian distribution model. It is therefore interesting to see whether the returns generated by our chaotic price model are fat-tail distributed. Fig. 17 shows the return distribution generated by the price dynamical model (1) with

---

[2] Notice that the drift $d(t)$ (26) is defined with sample averages, i.e. the averages are over the different Monte Carlo realizations of the price dynamical model (1) at the same time points, whereas the average $E_t\{*\}$ in the auto-correlation $E_t\{r_t r_{t+i}\}$ is computed over a single realization of the price dynamical model (1).

$(m,n)=(1,5)$, $w = 0.01$ and $a_1 = 0.14$, where a price trajectory of $10^5$ points was used to construct the distribution curve, and also shown in the figure is the Gaussian distribution (dashed line) with the same variance as the model (1) price returns. Comparing the two curves in Fig. 17 we see very clearly that the return distribution of model (1) is fat-tailed.

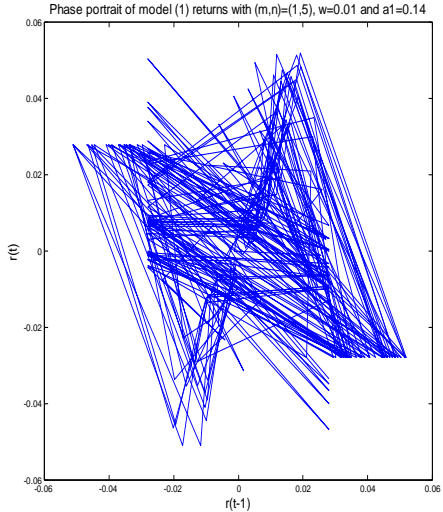

Fig. 16: Strange attractor: Phase portrait of model (1) returns on the 2D $r_{t-1}$-vs-$r_t$ subspace with $(m,n)=(1,5)$, $w = 0.01$ and $a_1(t) = 0.14$.

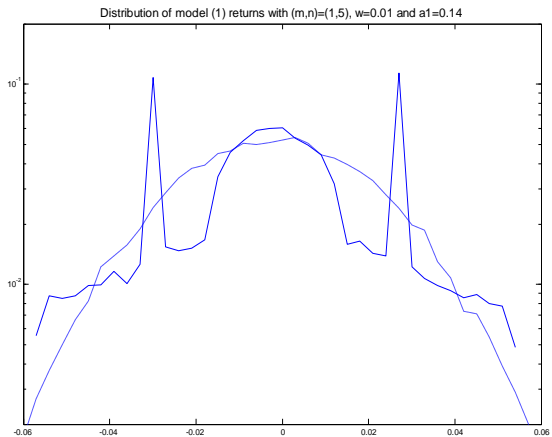

Fig. 17: Return distribution of the price dynamical model (1) in semi-log scale with $(m,n)=(1,5)$, $w = 0.01$ and $a_1 = 0.14$; the dashed line is Gaussian distribution with the same variance.

## IX. CONCLUDING REMARKS

The technical-trading-rule-based deterministic price dynamical models developed in this paper provide us a useful framework to analyze some key properties of stock prices and to view a number of important issues in financial economics from a different angle:

First, the classical concept of equilibrium and stability (in the Lyapunov sense) developed for Natural Systems may not be suitable for Social Systems (see [36] for the history of how Economics borrowed the concept of equilibrium from Physics). Specifically, as illustrated in Fig. 18, isolated equilibriums are common in Natural Systems and the classical equilibrium and stability concepts were developed for these scenarios; however, for Price Systems, the price will stay at any value forever if no buying or selling actions take place. In fact, the Price System illustrated in Fig. 18 is a schematic interpretation of Theorems 1 and 2 and the discussions in Section III: (i) any price is an equilibrium because the "price ball" will stay at any point if supply equals demand; (ii) all the equilibriums are unstable because a temporal small imbalance of supply and demand will move the "price ball" away from the price point and no natural force will push it back automatically (in contrast to the stable equilibriums of Natural System); and (iii) the Price System as a whole is quite stable in general --- the "price ball" is moving around from one point to another (regularly or chaotically) to digest the imbalance of supply and demand [8]. For Social Systems (in general and financial systems in particular), "moving around chaotically" is stable status, whereas "all moving in one direction" is the source of instability [4], [23], [37]. Hence, we need some new concepts of stability for Social Systems [25]; the concept of set-stable proposed in Section III is a trial in this regard.

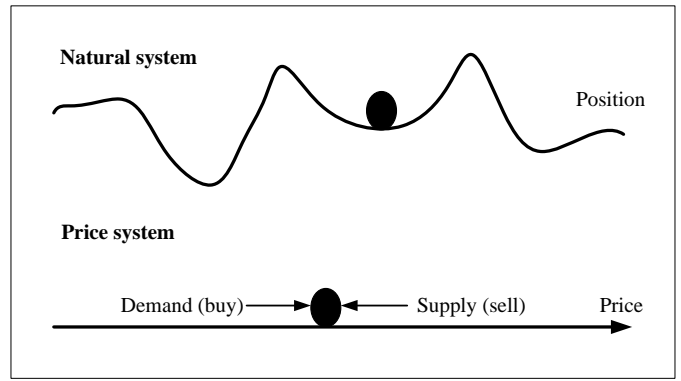

Fig. 18: Natural and Social (Price) Systems need different stability concepts.

Second, volatility is fixed function of model parameters which have clear physical meanings such as the strength of the traders ($a_1$), the magnitude of price rise (decline) around which the contrarians begin to act ($2w$), or the lengths of the price moving averages used in the technical trading rules ($m, n$). Consequently, the origins of the stylized facts about volatility such as volatility clustering and excessive volatility can be clearly identified: volatility clustering is due to the persistent actions of the traders who use their pre-determined strategy to buy or sell the stocks within a time interval until their objectives are achieved; and, excessive volatility is due to the strong actions of the traders within a very short period of time such as the pump-and-dump operations of the manipulators or the chain reactions of the stop-loss orders [26]. The insight, as illustrated in Fig. 19, is that volatility and price dynamical model are in the higher deterministic level which is slow time-varying [30] (the changes in model structure and parameters are slow in general as compared with the changes of prices), whereas price and return are in the lower chaotic or random level which is fast time-varying; that is, volatility should be viewed as a deterministic variable[3] and treated in the same way as model parameters.

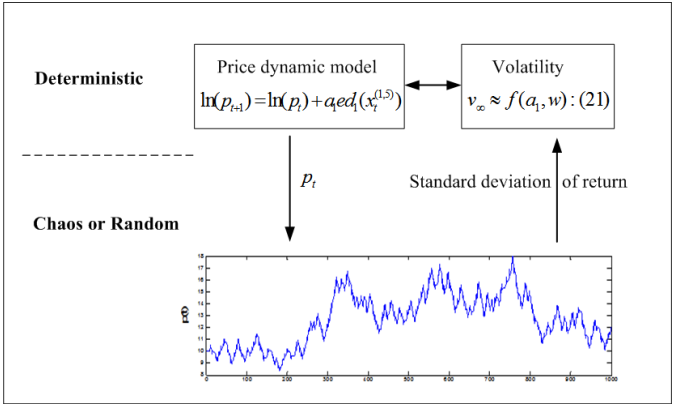

Fig. 19: Volatility is a deterministic variable and has a fixed relationship with the model parameters.

Third, short-term prediction is possible because the price dynamical model is purely deterministic, and the "prediction horizon" is characterized by the Lyapunov exponent which is a fixed function (20) of the model parameters. For the price dynamical model (1) with parameters in the typical chaotic range (see Figs. 3 and 11), it takes roughly two to six steps for the volatility to increase ten times (see Fig. 9); that is, suppose at a time point the strength parameter $a_1$ suddenly increases ten times (e.g. a big buyer starts to act at this time point), then it has a two to six time-step delay for the volatility to fully catch up with this change of trading activity.

Finally, uncorrelated returns (in the sense that the drift (26) equals the random walk drift $\sigma\sqrt{t}$) occur at some particular parameter values which are located at the central part of the chaotic zone. For parameters in other parts of the chaotic zone, the drift is either smaller than the random walk drift $\sigma\sqrt{t}$ (sub-diffusion) or larger than $\sigma\sqrt{t}$ (super-diffusion). The curves in Figs. 13 and 14 give us a clear picture of how the returns change from positively correlated (super-diffusion) to uncorrelated (diffusion) and then to negatively correlated (sub-diffusion) as the model parameters change. In this regard our deterministic price dynamical model provides a much richer framework than the random walk model to reveal the origin of return correlations.

---

[3] This is in contrast to the prevalent models such as ARCH [18], GARCH [7] and many others [2] that treat volatility as a random process driven by the same random source for the prices.

APPENDIX

**Proof of Lemma 1:** We consider the more general case of $m=1$ and $n$, $w$ and $a_1$ are free to change. Let the initial condition be $p_{-n+1} = \cdots = p_{-1} = p^*$ and $\ln(p_0) = \ln(p^*) + r_0$, i.e., the price stays at some arbitrary equilibrium price $p^*$ (any positive number) before time zero and at $t = 0$ there is a small price disturbance $r_0$ causing $p_0 = p^* e^{r_0}$. The basic idea of the proof is to compute $L_1 = \ln(r_1/r_0)$, $L_2 = \ln(r_2/r_1)$, $L_3 = \ln(r_3/r_2)$, …, until $Li \approx Li+1$ so that $Li$ is a good estimate of the Lyapunov exponent. From (2) and using $1 + s \approx e^s$ for small $s$ we have at $t = 0$ that

$$x_0^{(1,n)} = \ln\left(\frac{p^* e^{r_0}}{\frac{1}{n}(p^* e^{r_0} + p^*(n-1))}\right) \approx \left(1 - \frac{1}{n}\right) r_0 \quad (A1)$$

For small $r_0$ such that $\left(1 - \frac{1}{n}\right) r_0 < w$, we have from (1), (4) and (A1) that

$$r_1 = a_1 \, ed_1(x_0^{(1,n)}) \approx \frac{0.1 a_1}{w}\left(1 - \frac{1}{n}\right) r_0 \quad (A2)$$

So we get our first candidate for the Lyapunov exponent:

$$L_1 = \ln(r_1/r_0) = \ln\left(\frac{0.1 a_1}{w}\left(1 - \frac{1}{n}\right)\right) \quad (A3)$$

For $t = 1$, $p_1 = p_0 e^{r_1} = p^* e^{(r_0 + r_1)}$,

$$x_1^{(1,n)} = \ln\left(\frac{p^* e^{(r_0+r_1)}}{\frac{1}{n}(p^* e^{(r_0+r_1)} + p^* e^{r_0} + p^*(n-2))}\right)$$

$$\approx \left(1 - \frac{1}{n}\right) r_1 + \left(1 - \frac{2}{n}\right) r_0 \quad (A4)$$

and

$$r_2 = \frac{0.1 a_1}{w} x_1^{(1,n)}$$

$$\approx \left(\frac{n-2}{n-1} + \frac{0.1 a_1}{w}\left(1 - \frac{1}{n}\right)\right) r_1 \quad (A5)$$

for small $r_0$ such that $x_1^{(1,n)} < w$. From (A5) we have

$$L_2 = \ln(r_2/r_1) = \ln\left(\frac{n-2}{n-1} + \frac{0.1 a_1}{w}\left(1 - \frac{1}{n}\right)\right) \quad (A6)$$

Setting $n=5$ and $w=0.01$ in (A3) and (A6) we have $L_1 = \ln(8a_1)$ and $L_2 = \ln\left(\frac{3}{4} + 8a_1\right)$. For some typical values of $a_1$ in the chaotic zone, $\ln(8a_1)$ and $\ln\left(\frac{3}{4} + 8a_1\right)$ are not very close to each other (e.g., for $a_1 = 0.17$, $\ln(8a_1) = 0.3075$ and $\ln\left(\frac{3}{4} + 8a_1\right) = 0.7467$), therefore we move on to $t = 2$. From $p_2 = p_1 e^{r_2} = p^* e^{(r_0 + r_1 + r_2)}$,

$$x_2^{(1,n)} = \ln\left(\frac{p^* e^{(r_0+r_1+r_2)}}{\frac{1}{n}\left(p^* e^{(r_0+r_1+r_2)} + p^* e^{(r_0+r_1)} + p^* e^{r_0} + p^*(n-3)\right)}\right)$$

$$\approx \left(1 - \frac{1}{n}\right) r_2 + \left(1 - \frac{2}{n}\right) r_1 + \left(1 - \frac{3}{n}\right) r_0 \quad (A7)$$

and

$$r_3 = \frac{0.1 a_1}{w} x_2^{(1,n)}$$

$$\approx \left(\frac{\left(\frac{n-3}{n-1}\right) + \frac{0.1 a_1}{w}\left(1 - \frac{2}{n}\right)}{\left(\frac{n-2}{n-1}\right) + \frac{0.1 a_1}{w}\left(1 - \frac{1}{n}\right)} + \frac{0.1 a_1}{w}\left(1 - \frac{1}{n}\right)\right) r_2 \quad (A8)$$

for small $r_0$ such that $x_2^{(1,n)} < w$, we have

$$L_3 = \ln\left(\frac{r_3}{r_2}\right)$$

$$= \ln\left(\frac{\left(\frac{n-3}{n-1}\right) + \frac{0.1 a_1}{w}\left(1 - \frac{2}{n}\right)}{\left(\frac{n-2}{n-1}\right) + \frac{0.1 a_1}{w}\left(1 - \frac{1}{n}\right)} + \frac{0.1 a_1}{w}\left(1 - \frac{1}{n}\right)\right) \quad (A9)$$

Putting $n=5$ and $w=0.01$ into (A9) yields $L_3 = \ln\left(8a_1 + \frac{2+24a_1}{3+32a_1}\right)$. Fig. A1 plots $L_1 = \ln(8a_1)$, $L_2 = \ln\left(\frac{3}{4} + 8a_1\right)$ and $L_3 = \ln(8a_1 + \frac{2+24a_1}{3+32a_1})$ for $a_1$ in the chaotic zone. We see from Fig. A1 that $L_2$ and $L_3$ are very close to each other, therefore we can use any one of them, say $L_2 = \ln\left(\frac{3}{4} + 8a_1\right)$, as the Lyapunov exponent; this gives (19).

For the general case, we have from (A6) and (A9) that

$$e^{L_3} - e^{L_2} = \frac{\left(\frac{n-3}{n-1}\right) + \frac{0.1 a_1}{w}\left(1 - \frac{2}{n}\right)}{\left(\frac{n-2}{n-1}\right) + \frac{0.1 a_1}{w}\left(1 - \frac{1}{n}\right)} - \frac{n-2}{n-1}$$

$$= \frac{-n}{n(n-1)(n-2) + \frac{0.1 a_1}{w}(n-1)^3} \quad (A10)$$

which is in the order of $O(n^{-2})$, while $e^{L_2} = \frac{n-2}{n-1} + \frac{0.1 a_1}{w}\left(1 - \frac{1}{n}\right)$ is in the order of $O(1)$. Since $n$ is the length of the price moving averages whose common values are 5, 10, 20, …, this gives the relative difference $O(n^{-2})/O(1)$ in the order around 1% which is small, and consequently we can use $L_2$ of (A6) as the Lyapunov exponent; this proves (20). ∎

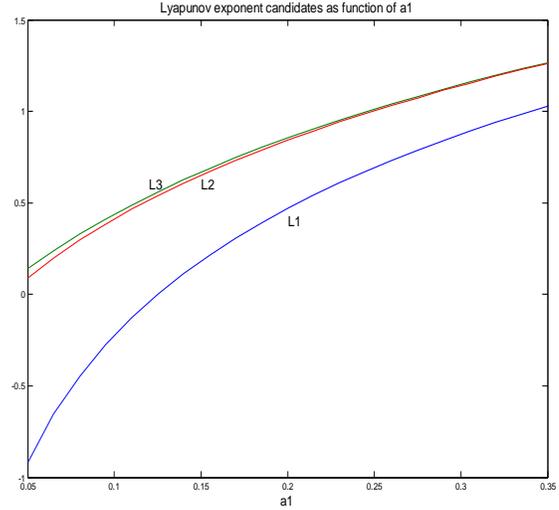

Fig. A1: Plots of the Lyapunov exponent candidates $L_1 = \ln(8a_1)$, $L_2 = \ln\left(\frac{3}{4} + 8a_1\right)$ and $L_3 = \ln(8a_1 + \frac{2+24a_1}{3+32a_1})$ as functions of the strength parameter $a_1$.

## ACKNOWLEDGMENT

The author would like thank the anonymous AE and the reviewers for their insightful comments that helped to improve the paper.

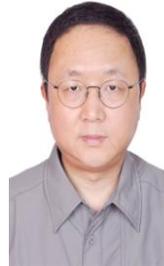

**Li-Xin Wang** received the Ph.D. degree from the Department of Electrical Engineering, University of Southern California (USC), Los Angeles, CA, USA, in 1992.

From 1992 to 1993, he was a Postdoctoral Fellow with the Department of Electrical Engineering and Computer Science, University of California at Berkeley. From 1993 to 2007, he was on the faculty of the Department of Electronic and Computer Engineering, The Hong Kong University of Science and Technology (HKUST). In 2007, he resigned from his tenured position at HKUST to become an independent researcher and investor in the stock and real estate markets in Hong Kong and China. In Fall 2013, he joined the faculty of the Department of Automation Science and Technology, Xian Jiaotong University, Xian, China, after a fruitful hunting journey across the wild land of investment to achieve financial freedom. His research interests are dynamical models of asset prices, market microstructure, trading strategies, fuzzy systems, and opinion networks.

Dr. Wang received USC's Phi Kappa Phi highest Student Recognition Award.